\documentclass[smallextended]{svjour3}       
\smartqed  

\usepackage{latexsym}
\usepackage{color}
\usepackage[usenames]{xcolor}
\usepackage{amssymb}
\usepackage{rotating}
\usepackage{multirow}
\usepackage{pdfsync}
\usepackage{url}

 \newif\ifdraft

\newcommand{\hasscomment}[1]{\ifdraft{\textcolor{brown}{[HS]: {#1}}}\else{\vspace{0ex}}\fi}
\newcommand{\walscomment}[1]{\ifdraft{\textcolor{green}{[WM]: {#1}}}\else{\vspace{0ex}}\fi}
\newcommand{\fabscomment}[1]{\ifdraft{\textcolor{red}{[FS]: {#1}}}\else{\vspace{0ex}}\fi}

\begin{document}

\title{Bridging Social Media via Distant Supervision
}

\titlerunning{Bridging Social Media via Distant Supervision}        

\author{Walid Magdy \and Hassan Sajjad \and Tarek El-Ganainy \and Fabrizio Sebastiani
}

\authorrunning{W. Magdy, H. Sajjad, T. El-Ganainy, F. Sebastiani} 

\institute{All the authors are at Qatar Computing Research Institute, Qatar Foundation, Doha, Qatar; \email{\{wmagdy,hsajjad,telganainy,fsebastiani\}@qf.org.qa}. Fabrizio Sebastiani is on leave from Consiglio Nazionale delle Ricerche, Italy.}

\date{Received: March 3, 2015}

\maketitle

\begin{abstract}
Microblog classification has received a lot of attention in recent years. Different classification tasks have been investigated, most of them focusing on classifying microblogs into a small number of classes (five or less) using a training set of manually annotated tweets. Unfortunately, labelling data is tedious and expensive, and finding tweets that cover all the classes of interest is not always straightforward, especially when some of the classes do not frequently arise in practice. In this paper we study an approach to tweet classification based on distant supervision, whereby we automatically transfer labels from one social medium to another for a single-label multi-class classification task. In particular, we apply YouTube video classes to tweets linking to these videos. This provides for free a virtually unlimited number of labelled instances that can be used as training data. The classification experiments we have run show that training a tweet classifier via these automatically labelled data achieves substantially better performance than training the same classifier with a limited amount of manually labelled data; this is advantageous, given that the automatically labelled data come at no cost. Further investigation of our approach shows its robustness when applied with different numbers of classes and across different languages. 
\end{abstract}

\section{Introduction}
\label{sec:intro}

\noindent Interest in classifying microblogs has increased with the widespread use of microblogging platforms such as Twitter. Tweets contain useful information that can be applied to various tasks, such as mass emergency management \cite{Imran:2014vn}, stock market analysis \cite{Bollen:2011bf}, social studies \cite{Dodds:2011hc}, and many others. Classifying tweets is usually an essential step in most such applications. From time to time, this may take the form of classification by topic, by sentiment, by political leaning, etc. One of the classification tasks that has received some (although still insufficient) attention is classifying tweets into general-purpose classes, such as e.g. \textsf{Politics}, \textsf{Sports}, \textsf{Entertainment}, \textsf{Science}, etc. Pre-classifying tweets under general-purpose classes can be useful in many applications, such as in online market research and advertising, social analysis of groups' or individuals' interests, and social search. \fabscomment{Can you guys insert a few citations here to support this statement?} In general, classifying tweets under general-purpose classes is an important enabling technology for applications that attempt to make sense of the Twitter firehose.

Classifying tweets involves several challenges. First of all, tweets contain a variety of information on a variety of topics, and given a specific tweet it is not easy to define an exact class for it. Consider the tweet 

\begin{quote}
\textit{The funniest reaction of a Barcelona supporter after the great goal by Messi \texttt{youtu.be/jLeTMIoAgCw}}
\end{quote} 

\noindent This tweet could be classified into classes such as \textsf{Sports}, \textsf{Comedy}, or \textsf{Entertainment}. As far as we know, there is no standard set of classes defined for microblogs that leverage the variety of information available on Twitter. Most microblog classes defined in previous works were motivated by specific applications \cite{1,2,4,7,10,14} and the number of classes was usually limited to a small number, typically five or less. The only work we are aware of that uses a substantively larger number (18) of microblog classes is \cite{11}; however, in this work the classes were derived from tweeting trends popular during a certain period, and these tend to change over time. 

A major challenge for standard classification approaches is the fact that manually annotated data are required to train an effective classifier. Data annotation is an expensive and time-consuming task, especially when a large number of classes is used, since a sufficient number of examples per class are required in order to yield reasonable classification accuracy. Sometimes, finding tweets that cover all the classes of interest is not straightforward, especially for classes that do not frequently arise in practice.

In this paper we present a novel method, based on distant supervision, for automatically deriving standard class labels for tweets, so as to generate a large number of training examples for microblog classification. Our proposed method does not require any manual annotation. We use crowdsourced labels from another social medium, YouTube, and we use these labels for training a tweet classifier. The benefits of using this method stem from the practically unlimited availability of training instances of this type. 
Furthermore, this method is language-independent; this makes it easy to train a classifier for tweets classification for any language available on Twitter.

We have collected a large set of tweets linking to YouTube videos. Each YouTube video is assigned to a class out of 18 predefined classes as a requirement when posting the video. We apply the classes assigned to the YouTube videos to the tweets which link them, which creates a large set of automatically labelled microblog data. We have then trained a classifier using hundreds of thousands of tweets linking to YouTube and covering 14 classes; in the literature, this is usually called \emph{distant supervision} \cite{go09}. We have then used this classifier to classify unlabelled tweets, and we have compared the results to those of a classifier trained using 1617 manually labelled tweets. The classifier trained via distant supervision turns out to yield substantially better classification accuracy than the one trained on manually annotated data.

We have analysed the effectiveness of our classification approach over different circumstances, so as to measure its robustness across different dimensions. We have first investigated the consequences of training our classifier with different sizes of automatically labelled data; here, we have found that training it with only 50,000 examples still outperforms the classifier trained with the manually labelled data. We have then run an additional experiment in which we have considered a smaller number of coarser classes (only 4, obtained by thematically grouping the original 14 classes); this experiment has shown that our classifier still outperforms the one trained with the manually labelled data. In an additional experiment we have compared the classification approaches on a test set of tweets dating from a time period much later than the one in which the training examples originated; the goal of this test was to investigate the effect of social media topical drift on classification effectiveness. In this last experiment our approach still achieved high performance, while a large drop in performance instead affected the classifier trained with the manually labelled data. Finally, we also tested our approach on a set of Arabic tweets, so as to study the language-independence of our distant supervision approach. Solid classification performance was noticed also on the Arabic test set.

The contributions of our study are thus the following:

\begin{enumerate}

\item Proposing a novel, efficient method for distant supervision by collecting automatically labelled data for the purpose of classifying microblogs under broad, general-purpose classes.

\item Proving that labels can be usefully transferred across different social media, thereby reducing the need of expensive manual labelling effort.

\item Investigating the effectiveness of our approach across several conditions including (1) different sizes of training examples, (2) different class granularity, (3) different degrees of recency, and 4) different languages.

\item Providing a set of 3,128 tweets manually labeled into 14 different classes, to be used as benchmark data for future research\footnote{The dataset is available for download at \url{http://alt.qcri.org/~wmagdy/resources.htm}.}.

\end{enumerate} 


\section{Related work}
\label{sec:relatedwork}

\noindent Distant supervision has been proposed in the literature for various applications, such as sentiment classification \cite{go09,marchettibowick-chambers:2012:EACL2012}, relation extraction \cite{mintz:acl09}, topical classification of blogs \cite{husby12}, and tweet classification \cite{zubiaga13}; this latter work will be discussed in detail later in this section. Most such works used distant supervision in order to obtain annotated data for their task from some other annotated dataset. For instance, \cite{go09} used the emoticons occurring in tweets as ``silver'' labels (i.e., as labels with more uncertain status that the ones found in usual ``gold'' standards) for tweet sentiment analysis. For relation extraction, \cite{mintz:acl09} used textual features extracted from Freebase relations in order to train a relation classifier; \cite{husby12} also used Freebase to obtain labels of Wikipedia articles, and used them for blog post classification by topic.

Previous work on microblog classification can be grouped according to three main dimensions: (1) the classification scheme used to classify the tweets, (2) the training method (e.g., standard supervision, distant supervision, etc.), and (3) the learning algorithm. As for (1), most published work for microblog classification focuses on classes targeted to a specific application. \cite{6} proposed a Wikipedia-based classification approach, by mapping tweets to the most similar Wikipedia pages; however, they tested their approach only on about 100 tweets grouped according to three events that occurred at the time of collecting the data. \cite{9} defined ten classes (e.g., \textsf{Musicians}, \textsf{Photography}, \textsf{Soccer}, \textsf{MartialArts}, \textsf{Motors}) for classifying blog posts. They used hyperlinks mentioned in the posts that link to webpages, and use the webpage metadata  for classifying the post. The metadata includes page title, description, tags, and categories, whenever any of them are available. They showed a substantial improvement in classification accuracy when using metadata information. Classifying tweets is however a more challenging task then classifying blog posts, because of the tweets' limited short sentence length. \cite{16} applied tweet-specific features in conjunction with bag-of-words to classify tweets into five broad classes (\textsf{News}, \textsf{Events}, \textsf{Opinions}, \textsf{Deals}, \textsf{PrivateMessages}). A simple classification task was discussed in \cite{14}, where tweets were classified as \textsf{News} or \textsf{Junk}; a similar work appeared in \cite{10}, where tweets linking to news articles were classified as \textsf{Comments} or \textsf{NewsReports}. Also in \cite{1} the authors performed binary tweet classification, discriminating \textsf{RealWorldEvents} from \textsf{NonRealWorldEvents}. \cite{7} and \cite{11} studied tweet classification over trending topics. \cite{11} is the only work we are aware of that uses a fairly comprehensive set of classes (18), thereby covering a vast portion of the Twitter-sphere. However, these classes were motivated from trending topics on Twitter, which tend to change over time.

Most previous work on tweet classification by topic uses manually annotated training data \cite{1,3,7,9,10,11,12,16}, which is both expensive and time-consuming. For training an effective classifier, a sizeable amount of training data is always required, especially when the number of classes is large. In addition, classifiers may need to be updated over time, so as to cope with concept drift, which may be especially severe in platforms as dynamic as those of social media. Therefore, methods that overcome the need for extensive manual annotation are to be preferred. \cite{3} applies a semi-supervised approach for classifying microblogs into six classes (which are a subset of the 14 classes used in our experimentation). They initially train a classifier with manually labelled data to probabilistically predict the classes for a large number of unlabelled tweets; then they train a new classifier also using the probabilistically predicted labels for the above-mentioned unlabelled tweets, and iterate the process to convergence.  \cite{zubiaga13} used distant supervision for tweet classification; as such, this work is highly relevant to the present work. Their approach consists in assuming that a tweet where a webpage URL occurs is on the same topic as that of the webpage; this is similar to our assumption about tweets mentioning YouTube links. The authors consider tweets linking to webpages classified under human-edited webpage directories. However, the 
shortcoming 
of their approach is that it depends on a human-edited directory which is limited in size and not necessarily up to date. 
Our proposed method is more robust, since it is not dependent on any manually maintained resource.

Regarding learning algorithms, different ones have been used in the literature for the tweet classification task, the most common being Na\"{\i}ve Bayes \cite{9,14,16}, decision trees \cite{7,11}, Labelled Latent Dirichlet Allocation (L-LDA) \cite{12}, and support vector machines (SVMs) \cite{10,11}.

Our work is different from the work reported in the literature in various respects. Considering the diversity of tweeted content, it is very hard to define classes for tweets that cover most of their aspects; we instead use standard classes from another social media platform. In addition, we propose a novel method for collecting automatically labelled data, to avoid the need for manually annotating training data. Our proposed method provides access to virtually unlimited amounts of free annotated data, amounts which can be increased at will essentially at no cost.


\section{Leveraging automatically obtained labels for microblog classification}


\subsection{Acquisition of automatically labelled tweets}
\label{sec:acquisition}

\noindent More than 4 million tweets in different languages linking to some YouTube video are tweeted everyday\footnote{\url{http://topsy.com/analytics?q1=site:youtube.com}}. Every video on YouTube is assigned one of 18 pre-defined classes at the time of its upload. Our approach for collecting labelled tweets is based on the hypothesis that a tweet linking to a YouTube video can be reasonably assigned the same class that the video has been assigned. To validate this hypothesis, we have assigned labels to tweets linking to YouTube videos and used them to train a tweet classifier.
We have used the Twitter API\footnote{\url{http://twitter4j.org/en/index.html}} with the string ``youtube lang:en'' to query the stream of English tweets with links to YouTube videos\footnote{This also captures tweets with shortened links to YouTube.}. We have thus collected a set of $\approx$ 19.5 million tweets with hyperlinks to $\approx$ 6.5 million different YouTube videos in a period of 40 days between the end of March and the beginning of May 2014; it is often the case that multiple tweets link to the same video. We have then used the YouTube API\footnote{\url{http://developers.google.com/youtube/}} to extract the titles and classes of these videos, and have assigned these video classes as labels to the tweets linking them. 

\begin{figure}
\centering
\includegraphics[width=\columnwidth]{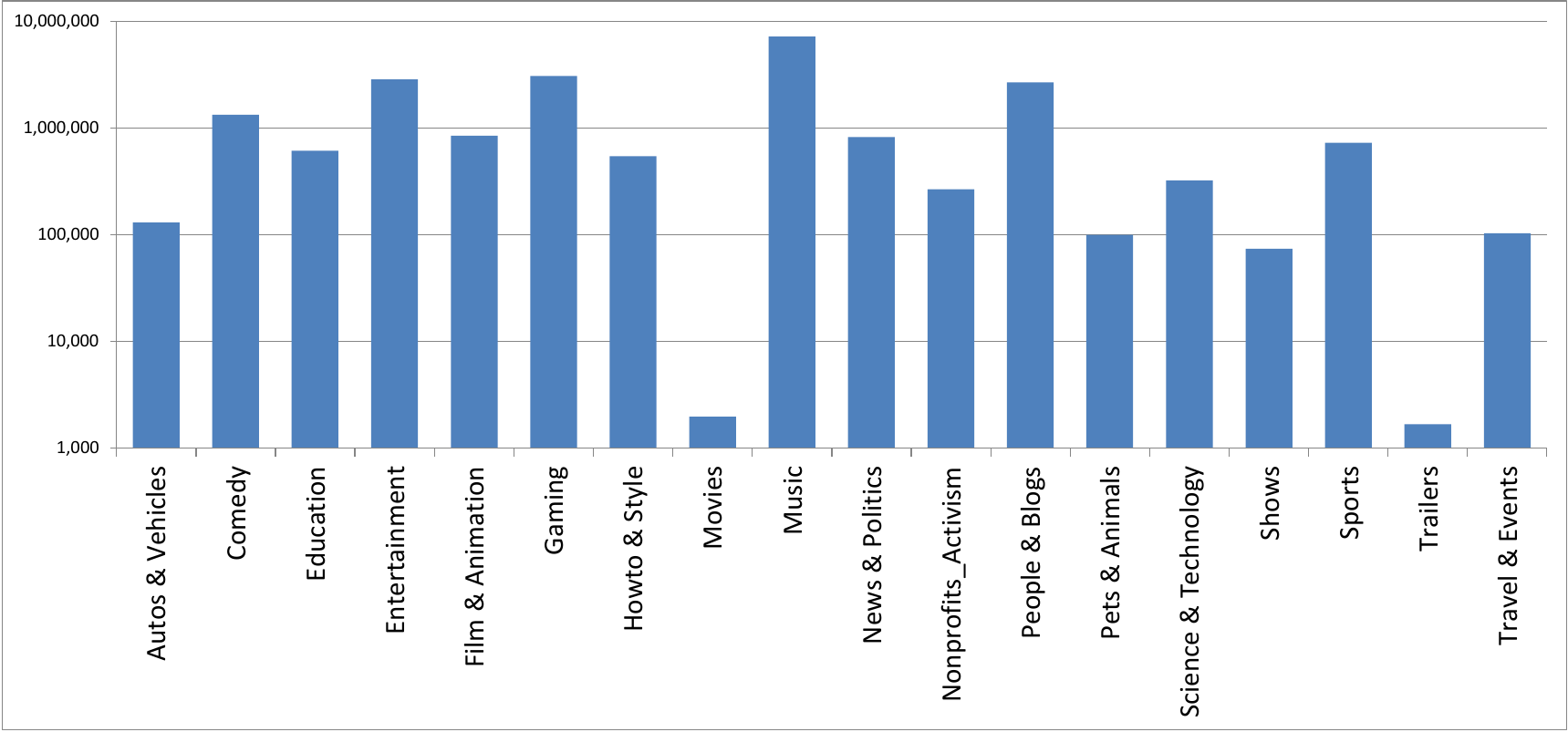}
\caption{Class distribution of the collected tweets.}
\label{dist}
\end{figure}

Figure \ref{dist} presents the distribution of the collected tweets across the 18 classes, plotted according to a log scale. As shown, the number of tweets per class ranges from only 1668 to more than 7 million. There are only three classes that contain less than 100k tweets (\textsf{Movies}, \textsf{Trailers}, and \textsf{Shows}). To avoid data sparseness, we have merged them with the class \textsf{Film\&Animation}, since these three classes are topically similar. \fabscomment{(NNTA) It's pretty peculiar to talk of ``data sparseness'' for classes that might have \emph{just} a few tens of thousand examples ...} The class \textsf{People\&Blogs} is the default class of YouTube, and is automatically assigned to videos when no class is specified by the user at the time of upload; we thus decided to drop this class, since we expect it to be noisy. \fabscomment{(NNTA) Well, but in this case you deprive yourself of a class that would be interesting for classifying tweets ...} Overall, these steps led to 14 classes with at least 100k tweets per class.

We have noticed that the collected tweet set contains large number of retweets and duplicate tweets, which are tweets with the same text. We have thus filtered out all the tweets that are retweets or have duplicate text, so as to keep at most one occurrence of each tweet in the dataset; this has the effect of avoiding to train the classifier with repeated examples, which may lead to bias. Moreover, duplicate tweets often contain automatically generated text (e.g., ``Just watched video ...''), which can act as noise when training the classification model. This step reduced our dataset size from $\approx$ 19.5 million to $\approx$ 9.2 million tweets only. In the end, the smallest class in our data contains $\approx$ 62k unique tweets.


\subsection{Features, feature selection, and model generation}
\label{sec:features}

\noindent In the tweet classification literature various types of features have been used for training a classifier. These include Twitter-specific features \cite{10,16}, social network features \cite{11}, hyperlink-based features \cite{9}, and standard bag-of-words features, which are the most commonly used \cite{1,6,7,11,14}. Since feature design is not our main focus in this paper we simply apply a bag-of-words (BOW) approach, where each feature represents a term and the feature value is binary, denoting presence or absence of the term in the tweet. Nonetheless, in the following we discuss two methods for text enrichment that attempt to improve the performance of the BOW approach.

Since the length of tweets is very short and the information contained in them is thus limited, we have applied two different feature enrichment methods in an attempt to improve classification accuracy. The first method enriches the tweet text in the training data with the title of the linked video. This method is only applicable to our automatically obtained training tweets, since they all link to YouTube, but is not applicable in general to the unlabelled tweets we want to classify, since these may not link to any YouTube video. The second method duplicates the hashtags contained in the tweets and removes the hash character ``\#'' from the second copy, so to allow the terms contained in the hashtags to increase the robustness of the term counts in the texts. 
In all our experiments, we applied simple text normalization, which includes case folding, elongation resolution (e.g., ``cooooool'' $\rightarrow$ ``cool''), and hyperlinks filtration. Neither stemming nor stop word removal were applied.

We have then applied feature selection, by scoring all features via information gain (IG), defined as
\begin{equation}\label{eq:IG}
 IG(t_{k}|c_{i})  =  H(c_{i})-H(c_{i}|t_{k})
  =  \sum_{c\in\{c_{i},\overline{c}_{i}\}}
 \sum_{t\in\{t_{k},\overline{t}_{k}\}} P(t,c) \log_{2}
 \frac{P(t,c)}{P(t) P(c)} \nonumber
\end{equation}
\noindent where $H(c_i)$ indicates the entropy of class $c_i$ and $H(c_i|t_k)$ indicates conditional entropy; probabilities are evaluated on the space of training documents, where $P(t_k)$ and $P((\overline{t}_k))$ represent the fractions of tweets that contain the term $t_k$ and do not contain $t_k$, respectively, and $P(c_i)$ and $P(\overline{c}_i)$ represent the fractions of tweets that are in class $c_i$ and are not in class ${c}_i$, respectively. $IG(t_k|c_i )$ measures the reduction in the entropy of $c_i$ obtained as a result of observing $t_k$, i.e., measures the information that $t_k$ provides on $c_i$. 

All features are ranked according to their IG value for the class, after which a round-robin mechanism \cite{5} is applied in which the top $n$ features are selected from each class-specific ranking and then merged to form the final feature space. In this way, for each class
$c_{i}$ the final set of selected features contains the $n$ features that are best at
discriminating $c_{i}$ from the other classes, which means that
all the classes in $C$ are adequately championed in
the final feature set.

We select the top 10,000 terms for each class; for 14 classes the theoretically maximum size of the feature space is thus 140,000 features, but the feature space is actually smaller since there is some overlap between the term sets selected for different classes.

After performing preliminary experiments with other learning algorithms (such as multinomial Na\"{\i}ve Bayes and the J48 decision-tree learner\footnote{\url{http://www.cs.waikato.ac.nz/ml/weka}}), we chose support vector machines (in Thorsten Joachims' SVM-light implementation \cite{8}) since they clearly appeared to outperform all others in both accuracy and efficiency.


\section{Experimental setup}
\label{sec:experiments}

\noindent In our experimental setup we have focused on testing the effectiveness of our method at classifying generic tweets, regardless of the fact that they link or not to a YouTube video. We created two test sets, 

\begin{itemize}

 \item an automatically labelled test set, harvested in the same manner as our training set (the ``silver standard''); and 

 \item a manually labelled test set, consisting of tweets that do not necessarily have links to YouTube videos (the ``gold standard'').

\end{itemize}


\subsection{Silver-standard training and test sets}
\label{sec:silverstandard}

\noindent From our dataset of automatically labelled tweets (described 
above) we randomly pick out for testing 1000 tweets for each class, for a total of 14,000 tweets evenly distributed across 14 classes. We refer to this test set as $test_{S}$ ($S$ standing for ``silver''). We consider $test_{S}$ as a ``silver standard'', since labels are not verified manually.
For the rest of the automatically labelled tweets, we have opted to balance the number of tweets in each class by randomly selecting 100,000 tweets from each class, so as to match the number of tweets in the smallest class, namely \textsf{Pets\&Animals}, which contains 98,855 tweets. The final training set thus contains $\approx$ 1.4 million tweets, which is three orders of magnitude larger than typical training sets used in the tweet classification literature. However, after applying duplicate and retweet filtering, as mentioned earlier, this number reduced to $\approx$ 913k tweets (each class having 60k to 70k examples). We refer to this dataset as $train_{S}$. We trained SVMs on $train_{S}$ using a linear kernel; this required a couple of hours on a standard desktop machine.


\subsection{Gold-standard training and test sets}
\label{sec:goldstandard}

\noindent We created a second test set (the ``gold standard'') consisting of manually labelled generic tweets; this test set will henceforth be referred to as $test_{G}$ ($G$ standing for ``gold''). There are two important reasons to have a manually labelled test set: 

\begin{enumerate}

 \item our $test_{S}$ silver standard may be biased in favour of the system trained on $train_{S}$ via distant supervision, because both datasets were sampled from the same distribution (i.e., they were labelled in the same automatic manner) and both consist of only tweets that link to YouTube; instead, the tweets in $test_{G}$ do not necessarily contain a link to a YouTube video;

 \item the manually labelled set $test_{G}$ can be used for cross-validation experiments, in the manner described below. This will provide a solid baseline for the classifier trained using $train_{S}$.

\end{enumerate}

\noindent To create a manually labelled set, it was difficult to randomly collect tweets covering all 14 classes, since some classes are rare and do not come up often in practice. In order to choose the tweets to label, we thus performed a guided search for each class by using the Twitter API to stream tweets that contain hashtags similar to class names. This was done in the same month in which we collected our automatically labelled training dataset. For example, for the class \textsf{Autos\&Vehicles} we collected tweets containing hashtags \#autos or \#vehicles. This helped us collect a set of tweets that, with high likelihood, had a substantial number of representatives for each of our classes of interest. \fabscomment{(NNTA) Doesn't this have the consequence that we end up in tailoring the test set to the classes we have chosen? E.g., assume that 90\% of the tweets in the Twittersphere are about nuclear waste disposal; none of them will end up in the test set, and in running our tests we will thus not realize that the set of classes is inadequate. In other words, the fact that the tweets that get manually tagged are not chosen at random biases the test set towards the set of classes we have chosen: ``If everything you have is a hammer, the world looks as made of nails.''} \hasscomment{This is a very good and dangerous point. We should discuss this!!} We randomly selected 200 tweets for each class (based on hashtags), removed the hashtags that relate them with their possible class, and submitted them to a crowdsourcing platform for annotation. For every tweet, we asked at least three annotators if the displayed tweet matches the assumed class or not.
Out of 2800 tweets representing 14 classes, only 1617 were assessed by all annotators as matching the assumed class; the number of tweets per class after validation ranged from 84 to 148. Examples of these tweets are shown in Table 3. This number of training examples is comparable to the numbers used in other studies from the literature \cite{1,6,7,10,11,14}.


\subsection{Classification runs}
\label{sec:classificationruns}

\noindent We have built the following classifiers for our experimentation:

\begin{itemize}

\item $C_{S}$: trained via distant supervision using $train_{S}$, which includes $\approx$ 913k automatically-labelled tweets.

\item $C_{S(v)}$: same as $C_{S}$, with tweet enrichment using the title of the linked \underline{v}ideo.

\item $C_{S(h)}$: same as $C_{S}$, with tweet enrichment obtained by adding the terms contained in the \underline{h}ashtags to the text.

\item $C_{S(vh)}$: same as $C_{S}$, with tweet enrichment obtained by both heuristics above.

\end{itemize}

\noindent The $S$ subscript indicates that all these classifiers have been trained on ``silver'' labels. 

Further to this, we have run 10-fold cross-validation (10FCV) experiments on the 1617 manually labelled tweets in $test_{G}$. We will then compare the results obtained by $C_{S}$ and its variants on $test_{G}$, with the ones obtained by the classifiers generated in these 10FCV experiments; specifically, we will look at the results of 

\begin{itemize}

\item $C_{G}$: this is not actually a single classifier but 10 different classifiers, as generated within the 10FCV; that is, the results of applying $C_{G}$ to $test_{G}$ will be the union of the 10 folds, each of them classified within one of the 10 experiments;

\item $C_{G(h)}$: similar to $C_{G}$, but with tweet enrichment obtained by adding the terms contained in the hashtags to the text. Enrichment using the title of the linked video is not applicable, since most of the tweets in $test_{G}$ do not link to YouTube.

\end{itemize}

\noindent Here, the $G$ subscript indicates that all these classifiers have been trained on ``gold'' labels.

The main objective of our experiments was to examine if any of the $C_{S}$ classifiers can achieve comparable (or even better) results with respect to the $C_{G}$ classifiers, which would support our hypothesis
and would also show the value of freely available labelled data. Different setups of the $C_{S}$ classifier were examined for both test sets to find the optimal configuration that achieves the best results.


\subsection{Evaluation}
\label{sec:evaluation}

\noindent The evaluation measures we used in this task are ``macroaveraged'' precision (\textbf{P}), recall (\textbf{R}), $F_{1}$ (popularly known as the ``F-measure''), and accuracy (\textbf{A}). That is, all of these measures were calculated for each class separately, after which the average was computed across the 14 classes. Since our test sets contain fairly balanced numbers of examples from each class 

\begin{itemize}

 \item these macroaveraged figures are very similar to the corresponding ``microaveraged'' ones (where classes more frequent in the test set weigh more), which are then not reported explicitly;
 
 \item accuracy is indeed a reasonable measure of classification effectiveness (this is unlike the cases of severe imbalance, when accuracy is unsuitable).

\end{itemize}


\section{Results}
\label{sec:results}

\noindent Table \ref{tab:resultSilver} and Table \ref{tab:resultGold} report the classification results obtained on the ``silver'' test set $test_{S}$ and on the ``golden'' test set $test_{G}$, respectively. All results in both tables display a relatively good effectiveness for a single-label 14-class classification task, where random classification would achieve (given the approximately balanced nature of our test sets) an expected classification accuracy of $\approx$ 7\%. 

Table \ref{tab:resultSilver} shows that the ``enhanced'' setups of the $C_{S}$ classifier did not lead to noticeable improvement. Enriching the training tweets with the title of the linked video even led to a small degradation in performance, while enriching the representation of the tweets by duplicating hashtags achieved only slightly better results. 

The results in Table \ref{tab:resultSilver} suggest that our idea of using YouTube labels for training a tweet classifier is a reasonable one. Nevertheless, the main experiments are those reported in Table \ref{tab:resultGold}, which reports results obtained on a truly gold standard.

\begin{table}
\centering
\small
\begin{tabular}{|l|cccc|}
\hline
 & \bf P & \bf R & $F_{1}$ & \bf A \\
\hline
$C_{S}$     & \bf 0.583 & 0.573 & 0.564 & 0.574 \\
$C_{S(v)}$  & 0.574 & 0.567 & 0.560 & 0.568 \\
$C_{S(h)}$  & 0.582 & \bf 0.575 & \bf 0.568 & \bf 0.576 \\
$C_{S(vh)}$ & 0.576 & 0.569 & 0.562 & 0.571 \\
\hline
\end{tabular}
\caption{\label{tab:resultSilver}Classification results on the silver-standard test set ($test_{S}$). \textbf{Boldface} indicates the best performer.}
\end{table}

\begin{table}
\centering
\small
\begin{tabular}{|l|cccc|}
\hline
 & \bf P & \bf R & $F_{1}$ & \bf A \\
\hline
$C_{G}$     & 0.511 & 0.506 & 0.507 & 0.518 \\
$C_{G(h)}$  & 0.541 & 0.534 & 0.537 & 0.546 \\
\hline
$C_{S}$     & \bf 0.619 & \bf 0.588 & \bf 0.579 & \bf 0.611 \\
$C_{S(v)}$  & 0.570 & 0.566 & 0.548 & 0.586 \\
$C_{S(h)}$  & 0.600 & 0.583 & 0.573 & 0.605 \\
$C_{S(vh)}$ & 0.578 & 0.567 & 0.551 & 0.588 \\
\hline
\end{tabular}
\caption{\label{tab:resultGold}Classification results on the gold-standard test set ($test_{G}$). \textbf{Boldface} indicates the best performer.}
\end{table}

Table \ref{tab:resultGold} reports the results of different setups of $C_{S}$ and $C_{G}$ on $test_{G}$. All different setups of $C_{S}$ achieved better performance than all different setups of $C_{G}$, which confirms that our method for inexpensively acquiring large numbers of automatically annotated training examples is more effective than the (more expensive) method of labelling a limited number of training examples.

Regarding the best setup for the training data, we noticed that hashtag term duplication improved the performance of $C_{G}$ over all scores, but did not lead to any improvement in the case of $C_{S}$. The limited number of training examples used for generating $C_{G}$ can be the reason for this result: here some enrichment to the representation of the training examples seems to help, unlike in the case of $C_{S}$, which was trained via a large number of training examples and does thus not require further enrichment. The best result achieved for $C_{S}$ and its variants was $A=0.611$ and $F_{1}=0.579$ (which was obtained for $C_{S}$ itself), which is substantially higher than the best result achieved for $C_{G}$ and its variants ($A=0.546$ and $F_{1}=0.537$, which was obtained for $C_{G(h)}$). From now on, $C_{S}$ and $C_{G(h)}$ will be used when comparing the distant supervision and the standard supervision approaches in further experiments. Anyway, the above result validates our hypothesis that classification labels from YouTube video could be applied to tweets linking them, and used to train a tweet classifier that is more effective than one obtained by manually labelling training data. 

Table \ref{tab:confusionMatrix-ut.mn} shows the complete confusion matrix obtained using the classifier $C_{S}$ on $test_{G}$. The classifier achieved a $F_{1}$ value higher than $0.650$ in classifying \textsf{Autos\&Vehicles}, \textsf{Gaming}, \textsf{HowTo\&Style}, \textsf{Pets\&Animals}, \textsf{Sports} and \textsf{Travel\&Events}. We further analysed the results of classes with a poor $F_{1}$ value. It is clear from the table that most of such classes were confused with the class \textsf{News\&Politics}. Other confusion between classes occurred, quite obviously, between related classes such as \textsf{Film\&Animation} and \textsf{Entertainment}, and \textsf{Comedy} and \textsf{Entertainment}. In some cases this may be due to the multifaceted nature of a tweet, that may naturally refer to more than one class. Examples of this phenomenon are the wrongly classified examples presented in Table \ref{tab:resultmisclassified}, where e.g., the tweet ``Thief Attacks Victim on Scooter'' is classified as \textsf{News\&Politics} instead of as \textsf{Autos\&Vehicles}. Both classes might be correct based on the content of the tweet. The examples presented in Table \ref{tab:confusionMatrix-ut.mn} show that some of the tweets can actually be classified into more than one class. This can motivate exploring multi-label multi-class classification in future work. \fabscomment{(NNTA) Essentially, we are here saying that tweet classification should probably be seen as a multi-label task, instead of single-label task as we are treating it. Perhaps we should have generated a multi-label classifier, even if the training examples are single-label (this can be done without problems, since you can indeed evaluate a multi-label classifier on a single-label test set).} \walscomment{Check final two sentences.} \fabscomment{Looks OK now.}


\begin{table*}[t]
 \resizebox{\textwidth}{!} {
\begin{tabular}{|r l | r r r r r r r r r r r r r r|}
\hline
 & & 1 & 2 & 3 & 4 & 5 & 6 & 7 & 8 & 9 & 10 & 11 & 12 & 13 & 14 \\ 
\hline
1 &\textsf{Autos\&Vehicles} & 136 & 0 & 0 & 1 & 0 & 0 & 0 & 1 & 4 & 0 & 1 & 2 & 1 & 2 \\ 
2 &\textsf{Comedy} & 3 & 33 & 1 & 7 & 8 & 3 & 3 & 8 & 2 & 2 & 11 & 4 & 9 & 7 \\ 
3 &\textsf{Education} & 0 & 0 & 23 & 0 & 1 & 2 & 4 & 2 & 20 & 3 & 0 & 33 & 5 & 1 \\ 
4 &\textsf{Entertainment} & 2 & 3 & 1 & 38 & 3 & 5 & 9 & 18 & 10 & 3 & 2 & 3 & 17 & 3 \\ 
5 &\textsf{Film\&Animation} & 4 & 1 & 4 & 5 & 55 & 2 & 3 & 5 & 8 & 2 & 3 & 6 & 5 & 13 \\ 
6 &\textsf{Gaming} & 1 & 1 & 0 & 2 & 3 & 105 & 3 & 3 & 3 & 0 & 4 & 12 & 6 & 1 \\ 
7 &\textsf{HowTo\&Style} & 4 & 1 & 0 & 3 & 4 & 5 & 86 & 2 & 0 & 1 & 9 & 4 & 5 & 1 \\ 
8 &\textsf{Music} & 1 & 0 & 1 & 1 & 5 & 0 & 4 & 55 & 2 & 2 & 1 & 2 & 5 & 5 \\ 
9 &\textsf{News\&Politics} & 8 & 1 & 2 & 2 & 1 & 1 & 2 & 7 & 56 & 1 & 2 & 10 & 4 & 5 \\ 
10 &\textsf{Nonprofits\&Activism} & 4 & 0 & 3 & 2 & 0 & 3 & 4 & 1 & 18 & 30 & 4 & 8 & 6 & 5 \\ 
11 &\textsf{Pets\&Animals} & 0 & 0 & 0 & 1 & 0 & 2 & 0 & 0 & 2 & 0 & 105 & 2 & 3 & 1 \\ 
12 &\textsf{Science\&Technology} & 2 & 0 & 3 & 1 & 2 & 4 & 3 & 2 & 16 & 3 & 1 & 69 & 3 & 4 \\ 
13 &\textsf{Sports} & 8 & 1 & 0 & 4 & 0 & 2 & 1 & 3 & 5 & 0 & 2 & 0 & 99 & 5 \\ 
14 &\textsf{Travel\&Events} & 4 & 0 & 1 & 0 & 2 & 1 & 8 & 5 & 6 & 1 & 4 & 6 & 3 & 98 \\ 
\hline
& Precision & 0.77 & 0.80 & 0.59 & 0.57 & 0.65 & 0.78 & 0.66 & 0.49 & 0.37 & 0.63 & 0.70 & 0.43 & 0.58 & 0.65 \\ 
& Recall & 0.92 & 0.33 & 0.24 & 0.32 & 0.47 & 0.73 & 0.69 & 0.65 & 0.55 & 0.34 & 0.91 & 0.61 & 0.76 & 0.71 \\ 
& $F_{1}$ & 0.84 & 0.46 & 0.35 & 0.41 & 0.55 & 0.75 & 0.67 & 0.56 & 0.44 & 0.44 & 0.79 & 0.50 & 0.66 & 0.68 \\ 
\hline
\end{tabular}
}
\caption{\label{tab:confusionMatrix-ut.mn} The confusion matrix for the classifier $C_{S(h)}$ as tested on $test_{G}$.}
\end{table*}

\begin{table*}[t]
\resizebox{\textwidth}{!} {
\begin{tabular}{|p{11cm}|l|l|}
\hline
\bf Tweet & \bf True Label & \bf Predicted Label \\
\hline
Female Softball Player Comes Out \#CelebrityNews \#Funny \#FunnyNews \#Jokes \url{http://t.co/K92JGuDArC} & \textsf{Comedy} & \textsf{Sports} \\
Thief Attacks Victim on Scooter & \textsf{Autos\&Vehicles} & \textsf{News\&Politics} \\
RT @Britt Coletti: State adopts new teacher & \textsf{Education} & \textsf{News\&Politics} \\
I learn \#German on my iPhone - just amazingly cool and only 99 cent \url{http://t.co/AwrsvfkLb8} \#ios \#cool & \textsf{Education} & \textsf{Science\&Technology} \\
\hline
\end{tabular}
}
\caption{\label{tab:resultmisclassified}A few examples of tweets misclassified by $C_{S(h)}$.}
\end{table*}


\section{Further experiments}
\label{sec:further}

\noindent In this section 

\begin{enumerate}

 \item we further investigate the robustness of our approach by measuring the consequences on classification effectiveness of increasing/decreasing the amount of training data;

 \item we examine the performance of classification using distant supervision when using a smaller number of coarser classes;

 \item we test how robust the classifiers trained on automatically labelled data are with respect to concept drift;

 \item we examine how language-independent our approach is by performing a classification experiment on non-English tweets (Arabic, in our case).

\end{enumerate}


\subsection{Effect of training data size on classification accuracy}
\label{sec:trainingdatasize}

\noindent In the previous section we have shown that, when compared on the same test set $test_{G}$, $C_{S}$ (the best of the classifiers trained via distant supervision, i.e., on silver labels) achieved substantially better results than $C_{G(h)}$ (the best of the classifiers trained on gold labels); this happened when using $\approx$ 913k training examples with $C_{S}$ vs.\ only 1617 for $C_{G(h)}$. Even though coming up with a dataset of $\approx$ 913k automatically labelled examples is much cheaper than coming up with one of 1617 manually annotated ones, it is interesting to study the effect of reducing the number of automatically annotated examples so as to see to what extent the automatically labelled data would retain its advantage. In addition, we have also examined the consequences of using more automatically annotated training examples, so as to see if there are further margins of improvement. 

\begin{figure}
\centering
\includegraphics[width=\columnwidth]{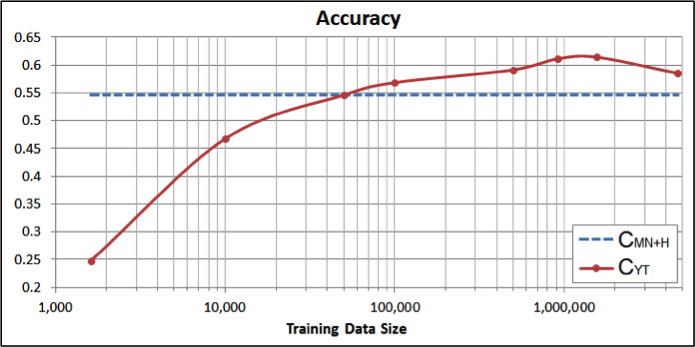}
\caption{Classification accuracy as a function of the amount of training data.}
\label{fig:graph}
\end{figure}

Figure \ref{fig:graph} shows a log-scale plot of classification accuracy as a function of the amount of silver training data. The dotted horizontal line represents the accuracy achieved by $C_{G(h)}$ using the 1617 manually labelled training examples. As shown, $C_{S}$ continues to outperform $C_{G(h)}$ when as few as $\approx$ 50k training examples are used; note that 50k tweets linking to YouTube videos covering all the classes could be easily collected in one day. However, with fewer than 50k automatically labelled training examples the performance of $C_{G(h)}$ is higher than that of $C_{S}$. When using the same small number of training examples (1617), the accuracy of $C_{S}$ is less than half the accuracy of $C_{G(h)}$. This highlights the fact that, as expected, YouTube-derived labels are not of the same quality as manually obtained ones. It thus makes sense, when using automatically derived labels, to use large numbers of them, especially since they come at essentially no cost.

We further tested the effects on classification accuracy of increasing the size of the training set even beyond 1.4m (which is the size of $train_{S}$ above); note however that this has the effect of disrupting the almost perfect balance among the classes, since (as previously mentioned) some classes had no more than 100k examples in our crawl. As shown in Figure \ref{fig:graph}, when increasing the size of the training data beyond 1.4m, accuracy slightly increased inasmuch as the imbalance was limited to the largest class having double the examples of the smallest class. However, when the level of imbalance went beyond that, accuracy suffered despite the larger size of the training set. 


\subsection{Testing distance supervision with smaller numbers of classes}
\label{sec:smallernumbersofclasses}

\noindent Our experiments so far had to do with classifying tweets into 14 mid-grained classes. A set of coarse classes can easily be extracted from the collected data. This increases the usability of the data for applications that require general classes.
\fabscomment{A quotation supporting this statement would also be in order here.}
\hasscomment{I changed the line completely and commented a sentence. please check!! I don't think it is critical to justify this experiment so we can simply remove these justifications or motivations}
To perform this, we have thematically grouped our 14 classes into only 4 classes. We have grouped \textsf{Education} with \textsf{Science\&Technology}, \textsf{News\&Politics} with \textsf{Nonprofits\&Activism}, \textsf{Autos\&Vehicles} with \textsf{Sports}, and the remaining classes \textsf{Comedy}, \textsf{Film\&Animation}, \textsf{Gaming}, \textsf{HowTo\&Style}, \textsf{Music}, \textsf{Pets\-\&Animals}, \textsf{Travel\&Events} with \textsf{Entertainment}. We have then retrained both $C_{G(h)}$ and $C_{S}$ using the new classification scheme. The results obtained on $test_{G}$ are shown in Table \ref{tab:result4classes}. As shown, $C_{S}$ continues to achieve superior performance with respect to $C_{G(h)}$, which further illustrates the effectiveness of distant supervision.

\begin{table}
\centering
\small
\begin{tabular}{|l|cccc|}
\hline
 & \bf P & \bf R & $F_{1}$ & \bf A \\
\hline
$C_{G(h)}$ & 0.593 & 0.588 & 0.590 & 0.699 \\
$C_{S}$    & \bf 0.710 & \bf 0.701 & \bf 0.705 & \bf 0.787 \\
\hline
\end{tabular}
\caption{\label{tab:result4classes}Classification results using 4 coarser classes instead of the 14 original ones. \textbf{Boldface} indicates the best performer.}
\end{table}


\subsection{Effects of concept drift on classification effectiveness}
\label{sec:timedrift}

\noindent One of the main characteristics of social media, and of Twitter in particular, is its highly dynamic nature, since the topics discussed change dramatically over time \cite{magdy2014adaptive}; as a consequence, the characteristics of tweets that belong to a certain class also tend to change, a phenomenon that in machine learning is called \emph{concept drift} \cite{Sammut:2011fk}. As a consequence, a model trained for a given tweet classification task could become less effective over time. In order to ascertain to what extent this problem affects our distant supervision method, we have carried out experiments in order to ascertain how much effectiveness drops when models trained by distant supervision are tested on tweets harvested several months after the models were trained. Most literature on tweet classification has so far neglected studying the consequences of concept drift. 

In December 2014 (i.e., 8 months after collecting all the data discussed in the previous sections) we have thus collected another set of tweets. Hashtags of class names were used to collect the tweets, then a random set of 200 tweets was selected from each class and annotated by crowdsourcers according to the same method used for creating $test_{G}$. Out of the 2800 tweets, only 1511 were assessed by the annotators to be matching the assumed class. We call the resulting test set $test_{G_{2}}$. We applied our two classifiers $C_{G(h)}$ and $C_{S}$ to the new test set $test_{G_{2}}$; results are reported in Table \ref{tab:result4time}. 

As shown in Table \ref{tab:result4time}, $C_{G(h)}$ suffered from a significant drop in performance, while $C_{S}$ obtained on $test_{G_{2}}$ results comparable to those obtained on $test_{G}$. This seems to suggest that one of the disadvantages of using a small number of manually labelled examples to train a tweet classification model is a drop in effectiveness over time due to the drift in social media content, which requires a robust model trained on a wide range of examples. Our findings point to another advantage of our distant supervision approach for tweet classification.

\begin{table}
\centering
\small
\begin{tabular}{|l|cccc|}
\hline
 & \bf P & \bf R & $F_{1}$ & \bf A \\
\hline
$C_{G(h)}$ & 0.462 & 0.456 & 0.450 & 0.465 \\
$C_{S}$    & \bf 0.615 & \bf 0.595 & \bf 0.587 & \bf 0.611 \\
\hline
\end{tabular}

\caption{\label{tab:result4time}Classification results on a test set of tweets collected 6 months later after the tweets used for training. \textbf{Boldface} indicates the best performer.}
\end{table}


\subsection{Experiments on non-English content}
\label{sec:arabic}

\noindent One of the main advantages of our approach is that it is language-independent, since no language-specific processing is required. Our final experiment thus concerned the application of our distant supervision approach to a language for which much fewer classification studies are available, i.e., Arabic. We collected a set of Arabic tweets linking to YouTube by running the query ``youtube lang:ar'' on the Twitter API. We collected more than 5 million tweets; the minimum number of tweets per class was 35,460 (for class \textsf{Pets\&Animals}). We extracted 1,000 tweets at random from each class for creating a ``silver'' test set, and selected from the remaining ones a balanced set of tweets to be used as ``silver'' training data. The final size of the training set was $\approx$ 482k, representing 14 classes. We attempted to use the same methodology of using hashtags for creating a manually labelled test set, but unfortunately the class names, once translated into Arabic, did not match enough tweets. Therefore, in this analysis we only rely on the ``silver'' test set only, which was shown in our earlier experimentation to be a good indicator of classification performance.

We applied one of the available tools for social Arabic text normalization \cite{darwish2012language}, which performs character normalization, word elongation resolution, and emotion detection. Normalized Arabic tweets were then used to train our classifier; as before, a BOW approach using IG for feature selection was used. 

Table \ref{tab:result4arabic} shows the results of classifying the Arabic tweet dataset. The results obtained are even higher than those on the English data, which illustrates the effectiveness of our distant supervision method regardless of the language in which tweets are expressed.

One interesting finding is that, as we noticed when extracting the meta-information from the videos linked by the Arabic tweets, 7\% of these videos have their title and description in a Latin-script language (mostly English). This shows that this approach could be applied even to languages with low resources on YouTube, since tweets in one language can link to videos titled in a different, resource-richer language.

\begin{table}
\centering
\small
\begin{tabular}{|l|cccc|}
\hline
 & \bf P & \bf R & $F_{1}$ & \bf A \\
\hline
$C_{S}$ & 0.644 & 0.646 & 0.640 & 0.646 \\
\hline
\end{tabular}
\caption{\label{tab:result4arabic}Classification results on the Arabic silver-standard test set.}
\end{table}


\section{Conclusion}
\label{sec:conclusion}

\noindent In this paper we have experimentally demonstrated the effectiveness of a ``distant supervision'' approach to tweet classification, consisting in automatically obtaining labelled data from one social media platform (YouTube) and using this data for training a classifier for another such platform (Twitter). Our proposed distant supervision method generates a large amount of freely available labelled training data, thus overcoming the need for manual annotations. As a side result, we have also generated a dataset of 3128 annotated tweets (the union of $test_{G}$ and $test_{G_{2}}$) that we make available to the research community.

When comparing the quality of a classifier trained via our distant supervision method with the one of a classifier trained on  $\approx$ 1.6k manually labelled tweets, we have shown that the former outperforms the latter when only $\approx$ 50k examples are used for training, which can be easily collected in one day using the freely available Twitter API. Our classification technique also showed superior effectiveness over the traditional one even when a smaller number of more general classes was considered instead. In addition we showed the robustness of our approach once used on resource-poor languages, and its robustness with respect to time drift.

For future work, it would be interesting to apply advanced pruning and data cleaning approaches for our collected training data, since it is collected automatically and is thus prone to noise; data cleaning could potentially  improve performance even further. In addition, it would be interesting to apply transfer learning \cite{13} in order to use our labelled tweets for different classification schemes. For example, this might be useful for sentiment analysis and emotion classification, since tweets of classes \textsf{Entertainment} and \textsf{Comedy} are more likely to be good indicators of positive emotions, while classes such as \textsf{News\&Politics} sadly tend to have  the opposite polarity.


\bibliographystyle{spmpsci}
\bibliography{DistantSupervision}

\end{document}